  \documentstyle[12pt]{article} 

  \topmargin - 0.5 cm 
\begin{document}

\baselineskip = 0.95 true cm


\vspace*{1.2 true cm}

\vspace{1.2 true cm}

\centerline{{\Large{\bf On the use of the group SO($4, 2$)}}} 

\centerline{{\Large{\bf in atomic and molecular physics}}}

\vspace{0.8 true cm}
\vspace{0.8 true cm}

\centerline{by M. R. KIBLER}

\centerline{Institut de Physique Nucl\'eaire de Lyon,}

\centerline{IN2P3-CNRS et Universit\'e Claude Bernard,}

\centerline{43 Boulevard du 11 Novembre 1918,}

\centerline{F-69622 Villeurbanne Cedex, France}

\vspace{6 true cm}

\centerline{to be published in Molecular Physics} 

\centerline{a Special Issue in Memory of Brian G. Wybourne}




\newpage 

\vspace*{1.2 true cm}

\centerline {\bf Abstract}

\bigskip

In this paper the dynamical noninvariance group SO($4, 2$) for a 
hydrogen-like atom is derived through two different approaches. The 
first one is by an established traditional ascent process starting 
from the symmetry group SO(3). This approach is presented in a 
mathematically oriented original way with a special emphasis on 
maximally superintegrable systems, $N$-dimensional extension and little 
groups. The second approach is by a new symmetry descent process 
starting from the noninvariance dynamical group Sp($8, {\bf R}$) for a 
four-dimensional harmonic oscillator. It is based on the little known 
concept of a Lie algebra under constraints and corresponds in some sense 
to a symmetry breaking mechanism. This paper ends with a brief discussion 
of the interest of SO($4, 2$) for a new group-theoretical approach 
to the periodic table of chemical elements. In this connection, a
general ongoing programme based on the use of a complete set of 
commuting operators is briefly described. It is believed that the
present paper could be useful not only to the atomic and molecular
community but also to people working in theoretical and mathematical 
physics.

  \newpage

\centerline{\normalsize {1. INTRODUCTION}}

\bigskip

Professor Brian Garner Wybourne held a central place in the theoretical studies 
of atomic, molecular and condensed matter physics. His untimely death in 
November 2003 in Toru\'n, Poland, came as a shock to the scientific community. 

Brian was an inspirational teacher who contributed significantly to the 
applications of symmetry in physics and chemical physics via the development 
of methods and models. He published part of 
his work on group theory and rare earth ions in crystals in three celebrated 
books [1-3]. He was active in many other fields 
(partly described in this 
Special Issue) either alone or in collaboration with many students and 
colleagues. The present author personally benefited enormously from Brian's 
work particularly in the following areas.

- Crystal-field theory (energy level calculation; relativistic crystal-field 
theory; intensity of electronic transitions between levels of partly-filled 
shell ions in solid-state or molecular environments).

- Group-theoretical methods in physics (through the use of characters, plethysms 
and Schur functions).

- Wigner-Racah calculus for finite or compact groups (especially for groups of 
interest in atomic and molecular physics).

A part of this paper in memory of Brian Wybourne  
could be regarded as an appendix to the chapter 
`Case Study II: The hydrogen atom' of the textbook 
`Classical Groups for Physicists' that Brian published in 1974 \cite{Wyb74-3}. One 
aim of the present author is to provide a bibliography of the 
many works on the quantum-mechanical treatment of the hydrogen atom that 
have appeared since 1974. The main body of the paper is devoted to the 
set up of the mathematical 
ingredients for an investigation of the periodic table in connection with dynamical 
groups for the Coulomb system and the harmonic oscillator system. 

This paper is organized in the following way. In \S~2, the 
familiar standard derivation (central to Wybourne's analysis) of the 
dynamical noninvariance group SO($4, 2$) for a hydrogen-like atom is  
revisited in a mathematical language in coherence with the new developments 
of \S~3. In \S~3, the derivation of the group SO($4, 2$) is obtained from a 
symmetry descent process rather than by a symmetry ascent process. The originality 
of the latter derivation is to be found in the use of the mathematical concept
of a Lie algebra under constraints developed by the present author and some of his 
collaborators. In the closing \S~4, the interest of the group SO($4, 2$) 
for a new group-theoretical description of the periodical charts for atoms, 
ions and molecules is touched upon and the general lines of an ongoing programme for 
a phenomenological although quantitative approach to the periodic table are given.

\bigskip
\bigskip

\centerline{\normalsize {2. FROM SO(3) TO SO($4, 2$)}}

\bigskip

A part of this section is devoted to a 
mathematical reformulation of known 
results contained in Wybourne's book [3] 
in order to prepare the ingredients 
necessary for \S~3 and \S~4. 

\bigskip

\centerline{2.1 The chain SO(3)~$\subset$~SO(4)}
  
\bigskip

Let
$$
H \Psi = E \Psi
\eqno (1)
$$
be the Schr\"odinger equation for a hydrogen-like atom of nuclear charge 
$Ze$ ($Z = 1$ for hydrogen). We recall that the discrete spectrum ($E < 0$) of 
the nonrelativistic Hamiltonian $H$ (in $N=3$ dimensions) is given by
$$
E \equiv E_3(n) = \frac{1}{n^2} \> E_3(1), \quad  \Psi \equiv \Psi_{nlm},
\eqno (2)
$$ 
where $E_3(1)$ (with $E_3(1) < 0$) is the energy of the ground state (depending on the reduced mass 
of the hydrogen-like system, the nuclear charge and the Planck constant) and 
where 
$$
n = 1, 2, 3, \cdots
$$
$$…
{\rm for \ fixed} \ n : l = 0, 1, \cdots, n - 1
$$
$$
{\rm for \ fixed} \ l : m = - l, - l+1, \cdots, l.
\eqno (3)
$$
The degeneracy degree of the level $E_n$ is 
$$
d_3(n) = n^2
\eqno (4)
$$ 
if the spin of the electron is not taken into account ($2n^2$ if taken into 
account).  

For fixed $n$ and fixed $l$, the degeneracy of the $2l+1$ wavefunctions 
$\Psi_{nlm}$ corresponding to different values of $m$ (for $l \not= 0$) 
is explained by the three-dimensional proper rotation group, 
isomorphic to SO(3). The degeneracy in $m$ follows from the existence of 
a first constant of the motion, i.e., the angular momentum 
${\bf L}(L_1,L_2,L_3)$ of the electron of the hydrogen-like atom. Indeed, 
we have a first set of three commuting operators: the Hamiltonian $H$, the 
square ${\bf L}^2$ of the angular momentum of the electron and the third 
component $L_3$ of the angular momentum.  The operators $L_1$, $L_2$ and 
$L_3$ span the Lie algebra of SO(3).

For fixed $n$, the degeneracy of the $n^2$ wavefunctions $\Psi_{nlm}$ corresponding 
to different values of $l$ (for $n \not= 1$) is accidental with respect to SO(3) 
and can be explained via the group SO(4), locally isomorphic to 
SU(2)$\otimes$SU(2). The introduction of SO(4) can be 
achieved in the framework either of a local approach credited to Pauli
\cite{Pau26-4} (see also \cite{Pau26-4+1})
or a stereographic approach due to Fock \cite{Foc35-5}. We shall 
follow here the local or Lie-like approach developed by Pauli \cite{Pau26-4}. The
degeneracy in $l$ (for $n \not= 1$) follows from the existence of a second 
constant of the motion, i.e., the Runge-Lenz vector ${\bf M}(M_1,M_2,M_3)$ 
(discussed by Laplace, Hamilton, Runge and Lenz in classical mechanics and by 
Pauli in quantum mechanics). By properly rescaling  the vector ${\bf M}$, we 
obtain a vector ${\bf A}(A_1,A_2,A_3)$ which has the dimension of an angular 
momentum. It can be shown that the set $\left\lbrace L_i, A_i : i = 1, 2, 3 \right\rbrace$
spans the Lie algebra of SO(4), SO($3, 1$) and E(3) for $E < 0$, $E > 0$ and 
$E = 0$, respectively. As a point of fact, the obtained Lie algebras are Lie 
algebras with constraints since there exists two quadratic relations between 
${\bf L}$ and 
${\bf A}$ \cite{Pau26-4}. We now continue with the discrete 
spectrum corresponding to the case $E < 0$. Then, by defining 
$$
{\bf J} = \frac{1}{2} ({\bf L} + {\bf A}), \quad  
{\bf K} = \frac{1}{2} ({\bf L} - {\bf A}),
\eqno (5)
$$
it is possible to write the Lie algebra of SO(4) in the form of the Lie algebra
of the direct product SO(3)$_J\otimes$SO(3)$_K$ with the sets 
$$
\left\lbrace J_i = \frac{1}{2} (L_i + A_i) : i = 1, 2, 3 \right\rbrace, \quad  
\left\lbrace K_i = \frac{1}{2} (L_i - A_i) : i = 1, 2, 3 \right\rbrace 
\eqno (6)
$$
generating SO(3)$_J$ and SO(3)$_K$, respectively. The unitary irreducible
representations (UIR's) of SO(4) can be labelled as ($j, k$) with 
$2j \in {\bf N}$ and 
$2k \in {\bf N}$. For fixed $n$, the $n^2$ functions $\Psi_{nlm}$ generate 
the UIR ($(n-1)/2, 
          (n-1)/2$) of SO(4), of dimension 
$n^2$, corresponding to $2j = 2k = n - 1$. The condition $j = k$ comes from 
one of the two quadratic relations between 
${\bf L}$ and 
${\bf A}$ (the
condition $j = k$ corresponds to 
${\bf J}^2 = 
 {\bf K}^2$ in terms of 
${\bf J}$ and 
${\bf K}$). The other relation gives $E = E_3(1)/n^2$. Finally, note that 
the restriction from SO(4) to SO(3) yields the decomposition 
$$
\left( \frac{n-1}{2}, \frac{n-1}{2} \right)
 = (0) \oplus (1) \oplus \cdots \oplus (n-1)
\eqno (7)
$$
in terms of UIR's ($l$) of SO(3). 

Similar developments can be achieved (along the line of the local approach) 
for the continuum spectrum corresponding to the case $E > 0$ and for the 
zero-energy point corresponding to the case $E=0$. This leads to the Lorentz
group in $3+1$ dimensions SO($3, 1$) for $E>0$ and to the Euclidean group in 
$N=3$ dimensions $E(3)$ for $E=0$. The two latter groups can also be obtained by
means of a stereographic projection from ${\bf R}^4$ to ${\bf R}^3$ (along the 
line of the global approach) (see Ref. \cite{Bar36-6} for SO($3, 1$)). 

At this point, it is interesting to understand the different status of the groups 
SO(3), from one side, and SO(4), SO($3, 1$) and E(3), from the other side, 
for a hydrogen-like atom. The special real orthogonal group SO(4), 
respectively SO(3), 
describes multiplets characterized by a given value 
of the principal quantum number $n$, 
respectively the orbital angular quantum number $l$; the multiplet of SO(4), 
respectively SO(3), associated with $n$, respectively $l$, is of dimension $n^2$, 
respectively $2l+1$. The compact group SO(3) is a geometrical symmetry group 
in the sense 
that it leaves invariant both the kinetic part and the potential part of the 
Hamiltonian $H$. As a consequence, the three generators $L_i$ ($i=1,2,3$) of SO(3) commute 
with $H$. Hence, we have the set $\left\lbrace H, {\bf L}^2, L_3 \right\rbrace$ of commuting 
operators. The compact group SO(4) 
is a dynamical invariance group that manifests itself here via its Lie algebra. The
set of the bound state vectors associated with an arbitrary energy level of $H$ 
spans a finite-dimensional UIR of SO(4). The group SO(4) 
does not correspond to a geometrical symmetry group of $H$ but its six generators 
$J_i$ and $K_i$ ($i=1,2,3$) commute with $H$. Hence, we have the set 
$\left\lbrace H, {\bf L}^2, L_3, {\bf A}^2, A_3 \right\rbrace$ of commuting 
operators. The latter set comprises five independent constants of the motion (the 
number $2N - 1 = 5$ is the maximum number that we can have for a dynamical 
system in $N=3$ dimensions) so that the hydrogen-like system is a maximally 
superintegrable system. Similarly, we may think of the noncompact 
groups SO($3, 1$) and E(3) as dynamical invariance groups.

Besides the symmetry group SO(3) and the three dynamical invariance groups SO(4), 
SO($3, 1$) and E(3), another group, namely SO($4, 2$),
the special real pseudo-orthogonal group in ${\bf R}^{4,2}$,
plays an important r\^ole. 
This group is locally isomorphic to SU($2, 2$), the special unitary group in 
${\bf C}^{2,2}$. It can be realized as a group of conformal transformations 
of the ordinary Minkowski space-time; this yields the conformal group 
(involving
three space rotations, three space-time rotations, four 
space-time translations, one dilatation and four 
special conformal transformations) that is the  
most general group leaving invariant the 
Maxwell equations. The group SO($4, 2$) 
was introduced, in 
connection with the hydrogen atom, by Malkin and Man'ko \cite{MalMan65-7},  
in a  SO($6, {\bf C}$) form, and independently by 
Barut and Kleinert \cite{BarKle68-8}
          (see also Refs.~\cite{BarKle68-8+1} and 
                          \cite{BarKle68-8+2}). The 
group SO($4, 2$)
is a dynamical noninvariance group in the sense that not all the generators 
of SO($4, 2$) commute with $H$. It can be considered as a spectrum generating 
group since any wavefunction       $\Psi_{nlm}$ can be deduced from any other 
wavefunction through the action on $\Psi_{nlm}$ of generators of SO($4, 2$) 
which do not commute with $H$. Indeed, the group SO($4, 2$) can serve for 
describing the complete spectrum (including the discrete case, the 
continuum case and the zero-energy case) of a hydrogen-like atom. The bound
state vectors and the continuum state vectors of a hydrogen-like atom can be 
classified by means of representations of the noncompact group SO($4, 2$). For 
example, the infinite set 
$$
\left\lbrace \Psi_{nlm} : n \in {\bf N}^*; l =  0,   1,  \cdots, n-1; 
                                      m = -l, -l+1, \cdots, l   \right\rbrace
\eqno (8)
$$ 
of bound state vectors generates an infinite-dimensional UIR, noted $h$, of
SO($4, 2$). In other words, we have
$$
h = (0, 0) \oplus (1/2, 1/2) \oplus (1, 1) \oplus \cdots. 
\eqno (9)
$$
This representation remains irreducible when restricting SO($4, 2$) to its
noncompact subgroup SO($4, 1$). Therefore, the group SO($4, 1$) is another 
dynamical noninvariance group for a hydrogen-like atom: the various bound 
state vectors of the Hamiltonian $H$ can be connected owing to shift operators 
of SO($4, 1$). Another relevant subgroup of SO($4, 2$) is the noncompact group 
SO($3, 2$). It is also a dynamical noninvariance group useful for describing 
the bound state and continuum state vectors of $H$. For instance, 
the set of the continuum state vectors spans an infinite-dimensional 
UIR of SO($3, 2$). In addition, the bound state vectors 
with $n+l$ even (respectively, $n+l$ odd) span an infinite-dimensional UIR 
$h_e$ (respectively, $h_o$) of SO($3, 2$) and the restriction from SO($4, 2$) to 
SO($3, 2$) yields 
$$
h = h_o \oplus h_e.
\eqno (10)
$$ 
The two de Sitter groups SO($3, 2$) 
and SO($4, 1$) can thus be considered as spectrum generating groups for a 
hydrogen-like system. 

Most of what preceeds can be extended from $N=3$ to $N$ arbitrary with 
$N \ge 2$ [12-14]. This leads to a hydrogen-like atom in $N$ dimensions described 
by a $N$-dimensional Schr\"odinger equation with an attractive $N$-dimensional 
Coulomb potential. The geometrical symmetry group for such a dynamical system 
is the rotation group in the space ${\bf R}^N$, a group isomorphic to SO($N$). The 
spectrum for the energy $E$ of this system has a discrete part 
($E < 0$), a continuous part ($E > 0$) and a zero-energy point ($E=0$). For the discrete 
part, the levels of 
energy $E_N(n)$ are characterized by a positive integer 
$n$. They are given by 
$$
E \equiv E_N(n) = \frac{(N-1)^2}{(2n + N - 3)^2} E_N(1), \quad n \in {\bf N}^*,
\eqno (11)
$$
where $E_N(1)$ (with $E_N(1) < 0$) is the energy of the ground level corresponding to $n=1$. The 
degree of degeneracy $d_N(n)$ of the energy level $E_N(n)$ is
$$
d_N(n) = \frac{(2n + N - 3)(n + N - 3)!}{(n-1)!(N - 1)!}
\eqno (12)
$$
(if the spin is not taken into consideration). For 
$n$ arbitrary in ${\bf N}^*$, the degeneracy of the $d_N(n)$ state vectors 
(corresponding to $E < 0$) associated with the level 
$E_N(n)$ follows from the dynamical invariance group 
SO($N+1$), a group isomorphic to the rotation group in the space
${\bf R}^{N+1}$. The group SO($N, 1$), a group isomorphic 
to a (spherical and hyperbolic) rotation group  
in the space ${\bf R}^{N,1}$,  
is a dynamical invariance group for the continuum
state vectors (corresponding to $E > 0$). Furthermore, the Euclidean group in $N$
dimensions E($N$), spanned by rotations and translations in the space 
${\bf R}^N$, is a dynamical invariance group for the zero-energy state
vectors (corresponding to $E = 0$). It should be noted that the groups
SO($N+1$), SO($N, 1$) and E($N$) are the so-called {\it little groups} of the Poincar\'e group in 
a Minkowski space in ($N+1, 1$) dimensions. Finally, when passing from 
$N=3$ to $N$ arbitrary, the dynamical noninvariance group SO($4, 2$) 
is  replaced  by SO($N+1, 2$). 

Going back to the case $N=3$, which is of central importance for 
this work, we now give the general lines for 
constructing SO($4, 2$) from SO(4) by an ascent process 
(cf.  Refs. \cite{Wyb74-3}, \cite{BarKle68-8}, \cite{Wul71-12} and \cite{Wul71-12+1}). The 
presentation adopted here is based on Jordan-Schwinger calculus 
and is adapted to the developments in \S~3. 

\bigskip

\centerline{2.2 The chain SO(4)~$\subset$~SO($4, 2$)}

\bigskip

The starting point is to find a bosonic realization of the Lie algebra of 
SO(4).  Since SO(4)~$\sim$~SU(2)$\otimes$SU(2)/Z$_2$,  we can use twice the 
Jordan-Schwinger bosonisation of SU(2) \cite{Sch65-24-13}. Let us introduce 
two commuting pairs of boson operators 
($a_1, a_2$) and ($a_3, a_4$). They satisfy the commutation relations
$$
[a_i, a_j^\dagger] = \delta(i, j), \quad 
[a_i, a_j        ] = [a_i^\dagger, a_j^\dagger] = 0,
\eqno (13)
$$
where the indices $i, j$ can take the values 1 to 4 and where we use 
$A^\dagger$ to denote the Hermitean conjugate of the operator $A$. The $a_i$'s 
are annihilation operators and the $a_i^\dagger$'s are creation operators. It 
is a simple matter of calculation to check that the six bilinear forms 
($J_{12}, J_{23}, J_{31}$) and ($J_{14}, J_{24}, J_{34}$) defined via 
$$
J_{ab} = \frac{1}{2} 
({\bf a}^\dagger \> \sigma_c \> {\bf a} + 
 {\bf b}^\dagger \> \sigma_c \> {\bf b}) \quad {\rm with} \quad a, b, c \quad {\rm cyclic}
\eqno (14{\rm a})
$$
and
$$
J_{a4} = - \frac{1}{2} 
({\bf a}^\dagger \> \sigma_a \> {\bf a} - 
 {\bf b}^\dagger \> \sigma_a \> {\bf b})
\eqno (14{\rm b})
$$
span the Lie algebra $D_2$ of SO(4). In the latter two definitions, we have

- the indices $a, b, c$ can take the values 1 to 3

- $\sigma_1$, $\sigma_2$ and $\sigma_3$ are the three Pauli matrices

- ${\bf a}$ and 
  ${\bf b}$ stand for the column vectors whose transposed 
vectors are the line vectors $^t{\bf a} = (a_1 a_2)$ and $^t{\bf b} = (a_3 a_4)$, 
respectively

- ${\bf a}^\dagger$ and 
  ${\bf b}^\dagger$ stand for the line vectors 
${\bf a}^\dagger = (a_1^\dagger a_2^\dagger)$ and 
${\bf b}^\dagger = (a_3^\dagger a_4^\dagger)$, respectively. 

We can find nine additional operators which together with the two triplets of 
operators ($J_{12}, J_{23}, J_{31}$) and 
          ($J_{14}, J_{24}, J_{34}$) generate the Lie algebra $D_3$ of 
SO($4, 2$). These generators can be defined as 
$$
J_{a5} =   i [J_{a4}, J_{45}] \quad {\rm with} \quad a = 1, 2, 3,
$$
$$
J_{45} = \frac{1}{2} ({\bf a}^\dagger \> \sigma_2 \> ^t{\bf b}^\dagger - 
                      {\bf a}         \> \sigma_2 \> ^t{\bf b}),
$$
$$		      
J_{a6} = - i [J_{a5}, J_{56}] \quad {\rm with} \quad a = 1, 2, 3,
$$
$$
J_{46} = - i [J_{45}, J_{56}],
$$
$$
J_{56} = \frac{1}{2} ({\bf a}^\dagger {\bf a} + 
                      {\bf b}^\dagger {\bf b} + 2).
\eqno (15)
$$
The fifteen operators ($J_{12}, J_{23}, J_{31}$), 
                      ($J_{14}, J_{24}, J_{34}$), on one hand, and 
($J_{15}, J_{25}, J_{35}, J_{45}$), 
($J_{16}, J_{26}, J_{36}, J_{46}, J_{56}$), on the other hand,  
can be shown to satisfy 
$$
[J_{ab}, J_{cd}] = i(g_{bc}J_{ad} - 
                     g_{ac}J_{bd} + 
		     g_{ad}J_{bc} - 
		     g_{bd}J_{ac})
\eqno (16)
$$
with the metric tensor ($g_{ab}$) defined by
$$
(g_{ab}) = {\rm diag}(-1, -1, -1, -1, +1, +1),
\eqno (17)
$$
where the indices $a, b, c, d$ can take the values 1 to 6. We thus end up with 
the Lie algebra of SO($4, 2$), a group which leaves the real quadratic 
form
$$
- \sum_{a, b} g_{ab} \> x_a 
                     \> x_b = x_1^2 + 
                              x_2^2 + 
			      x_3^2 + 
			      x_4^2 - 
			      x_5^2 - 
			      x_6^2
\eqno (18)
$$
invariant. A set of infinitesimal generators 
of SO($4, 2$), acting on functions 
$f : (x_1, x_2, \cdots, x_6) \mapsto 
    f(x_1, x_2, \cdots, x_6)$ of six variables, 
is provided by the differential operators 
$$
J_{ab} \mapsto U_{ab} = i \left( g_{aa} \> x_a \> \frac{\partial}{\partial x_b} - 
                                 g_{bb} \> x_b \> \frac{\partial}{\partial x_a} \right)
\eqno (19)
$$
which generalize the components of the usual orbital angular momentum in the ordinary 
space ${\bf R}^3$. 

Since the Lie algebra of SO($4, 2$) is of rank 3, we have three invariant operators in 
its enveloping algebra. Indeed, they are of degree 2, 3 and 4  and are given by
$$
C_1 = \sum J_{ab} J^{ab},
$$
$$
C_2 = \sum {\epsilon}_{abcdef} J^{ab} J^{cd} J^{ef},
$$
$$
C_3 = \sum J_{ab} J^{bc} J_{cd} J^{da},
\eqno (20)
$$
where ${\epsilon}$ is the totally anti-symmetric tensor on six covariant indices
with ${\epsilon}_{123456} = 1$ and 
$$
J^{ab} = \sum g^{ac} \> g^{bd} \> J_{cd} \quad {\rm with} \quad g^{ab} = g_{ab},
\eqno (21)
$$
where we use the Einstein summation conventions.

We now briefly discuss the existence of the 
special representation $h$ of SO($4, 2$) that 
provides the quantum numbers $n$, $l$ and 
$m$. It can be seen that the operator 
$J_{45}$ connects states having different 
values of $n$. This clearly shows that the 
Lie algebra of the group SO($4, 2$) is a dynamical 
noninvariance algebra. In fact, it 
is possible to construct any state vector 
$\Psi_{nlm}$ by repeated application on the ground 
state vector $\Psi_{100}$ of the operator 
$J_{45} - J_{46}$ and of shift operators 
for SO(4). This result is at the root of the fact 
that all discrete levels (or bound 
states) of a hydrogen-like atom span an UIR of 
SO($4, 2$). This is the representation 
$h$ discussed above. It corresponds to the 
eigenvalues 6, 0 and -12 of the invariant 
operators $C_1$, $C_2$ and $C_3$, respectively. The 
dynamical noninvariance group 
SO($4, 2$) contains the dynamical noninvariance 
group SO($4, 1$) that 
contains in turn the dynamical invariance group 
SO(4) and thus the geometrical symmetry group 
SO(3). The restriction of 
SO($4, 2$) to SO(4) yields the decomposition
$$
h = \bigoplus_{n \in {\bf N}^*} \> \left( \frac{n-1}{2}, \frac{n-1}{2} \right)
\eqno (22)
$$
which is a direct sum of the representations ($j, j$), with $2j \in {\bf N}$, of the 
group SO(4). Further restriction from SO(4) to SO(3) gives
$$
h = \bigoplus_{n = 1}^{\infty} \bigoplus_{l=0}^{n-1} \> (l)
\eqno (23)
$$
that reflects that all the discrete states (with $E < 0$) of a hydrogen-like atom are 
contained in a single UIR of SO($4, 2$). A similar result holds for the continuum states 
(with $E > 0$). 

\bigskip
\bigskip

\centerline{\normalsize {3. FROM Sp($8, {\bf R}$) TO SO($4, 2$)}}

\bigskip

\centerline{3.1 Two important dynamical systems}

\bigskip

The hydrogen atom and the harmonic oscillator are two dynamical systems of
paramount importance for nuclear, atomic and molecular physics. The harmonic
oscillator turns out to be a corner-stone in nuclear spectroscopy, molecular
dynamics and the theory of radiation while the hydrogen atom is of considerable
importance in the theory of atomic and molecular structure. They can be 
considered as two paradigms among the 
exactly solvable dynamical systems. In fact, hydrogen-like atoms
are the sole atoms for which we know how to solve exactly the Schr\"odinger
equation. (In passing, let us
remember that hydrogen constitutes the 3/4 of the known universe.) From the
point of view of classical mechanics, each of these two systems (in $N=3$ 
dimensions) corresponds to the motion of a particle in an attractive spherically
symmetric potential (the Coulomb potential and the isotropic harmonic oscillator
potential) for which the bounded trajectories are closed \cite{Ber73-14} (the
case of the hydrogen atom being associated with the Kepler motion). In addition,
the two systems (both from a classical and a quantum-mechanical aspect) are
maximally superintegrable systems with $2N-1=5$ constants of the motion (see for
example Ref. \cite{KibWin90-15}). 

A connection between the hydrogen atom and the isotropic harmonic oscillator has
been recognized for a long time by numerous authors [20-42]. Such a connection 
can be established via Lie-like methods (local or infinitesimal approach) 
or algebraic methods
based on nonbijective canonical transformations (global or 
partial differential equation approach). From these
two approaches, to be described below, it is possible to derive the dynamical
noninvariance group SO($4, 2$) for a hydrogen-like atom in $N=3$ dimensions from 
the dynamical noninvariance group Sp($8, {\bf R}$) for an isotropic harmonic
oscillator in $N=4$ dimensions. This will be done in \S~3.4 in an original way.

\bigskip

\centerline{3.2 The global approach}

\bigskip

It is based on the Kustaanheimo-Stiefel transformation introduced in celestial
mechanics for regularizing the Kepler problem [22, 23]. This transformation is the 
${\bf R}^4 \to {\bf R}^3 : (u_1, u_2, u_3, u_4) \mapsto (x_1, x_2, x_3)$ 
surjection defined by 
$$
x_1 = 2 (u_1u_3 - u_2u_4),
$$
$$
x_2 = 2 (u_1u_4 + u_2u_3),
$$
$$
x_3 = u_1^2 + u_2^2 - u_3^2 - u_4^2),
\eqno (24{\rm a})
$$
accompanied by the constraint
$$
- u_1 du_2 + u_2 du_1 + u_3 du_4 - u_4 du_3 = 0.
\eqno (24{\rm b})
$$
This nonbijective canonical transformation can be 
identified with the Hopf fibration 
${\rm S}^3 \times {\bf R}_+ \to 
 {\rm S}^2 \times {\bf R}_+$ with compact fiber 
${\rm S}^1$ and can be introduced in 
a natural way through the theory of spinors [39-41]. 

The application of the Kustaanheimo-Stiefel transformation 
to the Schr\"odinger equation $H \Psi = E \Psi$ for a hydrogen-like atom 
(in $N=3$ dimensions) leads to a Schr\"odinger equation for an isotropic
harmonic oscillator in $N=4$ dimensions subjected 
to a constraint [27-30, 32-38]. In Ref. [35], it 
was shown that the Schr\"odinger equation for the oscillator is amenable 
to a system of Schr\"odinger equations for 

- a pair of two-dimensional isotropic harmonic oscillators 
with attractive potentials and subjected to a constraint in the case $E < 0$,

- a pair of two-dimensional isotropic harmonic oscillators 
with repulsive potentials and subjected to a constraint in the case $E > 0$,

- a pair of two-dimensional free particle systems  
subjected to a constraint in the case $E = 0$. 

\bigskip

\centerline{3.3 The local  approach}

\bigskip

The result just enunciated can be also obtained by combining the Pauli 
approach \cite{Pau26-4} of a hydrogen-like atom with the Schwinger approach 
\cite{Sch65-24-13} of angular momentum. By way of illustration, we consider the
case $E < 0$. In that case, the noninvariance dynamical algebra of the Lie group
SO(4) can be converted into the Lie algebra of the group SU(2)$\otimes$SU(2) 
with a constraint. According to Schwinger \cite{Sch65-24-13}, each group SU(2) 
which describes a generalized angular momentum can be associated with a
two-dimensional isotropic harmonic oscillator. We are thus left with the
following result: The discrete spectrum (corresponding to $E < 0$) for a
hydrogen-like atom can be deduced from the quantization of a pair of 
three-dimensional generalized
angular momenta with constraint or alternatively a pair of two-dimensional 
isotropic harmonic oscillators with attactive potentials 
and having the same energy. Similar results hold for $E > 0$ 
and $E = 0$.

\bigskip

\centerline{3.4 Connecting the dynamical noninvariance algebras}

\bigskip

We are now in a position to derive in a very simple way the 
dynamical noninvariance group SO($4, 2$) for a hydrogen-like 
atom. 

At this stage, we have to recall that the dynamical 
noninvariance group for the isotropic harmonic oscillator 
in $N$ dimemensions ($N \ge 1$) is the real symplectic 
group Sp($2N, {\bf R}$) [43-49]. This noncompact group
admits the group SU($N$) as a maximal compact subgroup. The
latter subgroup is the dynamical invariance group for the 
$N$-dimensional oscillator with attractive potential [43-45]. This 
dynamical system presents the energy levels
$$
F_N(n) = \frac{2n + N}{N} F_N(0), \quad n \in {\bf N}, 
\eqno (25)
$$
where $F_N(0)$ (with $F_N(0) > 0$) stands for the energy of the ground level 
corresponding to $n=0$. The degree of degeneracy $e_N(n)$ of the energy level 
$F_N(n)$ is
$$
e_N(n) = \frac{(n+N-1)!}{n! (N-1)!}. 
\eqno (26)
$$

We are interested here in the situation where 
$N=4$ and therefore with the group 
Sp($8, {\bf R}$)
as the dynamical noninvariance group for a four-dimensional 
isotropic harmonic oscillator. As a remarkable result, the 
constraint for $E < 0$, $E > 0$ and $E = 0$ assumes the same form 
and amounts to take as zero one of the thirty-six 
generators of Sp($8, {\bf R}$). This produces a Lie algebra under 
constraints that is isomorphic to the Lie algebra of SO($4, 2$). An
original derivation of central importance for the future developments 
of \S~4 is in order.

The constraint Cartan 1-form defined by Eq.~(24b) gives rise to a vector field 
which can be seen to belong to the Lie algebra of Sp($8, {\bf R}$). This vector
field generates a one-dimensional Lie algebra isomorphic to the Lie subalgebra 
of a subgroup SO(2) of Sp($8, {\bf R}$). Then, the dynamical noninvariance group 
for the hydrogen-like results from the vanishing of the vector field under 
consideration inside the Lie algebra of Sp($8, {\bf R}$). This produces a Lie 
algebra under constraints that can be derived as follows. According to the concept 
of Lie algebra under constraints \cite{KibWin88-26}, we have to find first the centraliser
of the Lie algebra of SO(2) into the Lie algebra of Sp($8, {\bf R}$). We obtain
$$
{\rm cent}_{{\rm sp}(8, {\bf R})} {\rm so(2)} = {\rm u(2, 2)}
\eqno (27)
$$
(we use lower case letters for Lie algebras). Then, the Lie algebra we are looking for 
is given by the factor algebra of this centralizer by so(2). We thus get
$$
{\rm cent}_{{\rm sp}(8, {\bf R})} {\rm so(2)} / {\rm so(2)} = 
{\rm su(2, 2)} \sim {\rm so(4, 2)}
\eqno (28)
$$
In simple terms, the Lie algebra so(4, 2) of the Lie group SO(4, 2) appears
as the Lie algebra that survives when one forces the generator of 
SO(2) to vanish within the Lie algebra of 
${\rm Sp}(8, {\bf R}) \supset {\rm SO}(2)$. 

As a conclusion, the dynamical noninvariance algebra of
the group SO($4, 2$) for a hydrogen-like atom in ${\bf R}^3$ 
was obtained from
the dynamical noninvariance algebra of the group Sp($8, {\bf R}$) 
for an isotropic harmonic oscillator in ${\bf R}^4$. This derivation 
of SO($4, 2$) corresponds to a symmetry 
descent process and should be contrasted with the one in \S~2.2 
which corresponds to a symmetry ascent process. A link between the two 
derivations can be established by noting that the Lie algebra of 
Sp($8, {\bf R}$) 
can be generated by the thirty-six possible bilinear forms 
constructed from the annihilation operators $a_i$ and the 
creation operators $a_i^{\dagger}$ 
($i = 1$ to $4$) used in \S~2.2 and that fifteen of these bilinear forms 
generate the Lie algebra of SU($2, 2$), the covering group of SO($4, 2$).
When the vector field that generates SO(2) is written in terms of the
annihilation and creation operators of Eq.~(13), the Lie algebra given by
Eq.~(28) appears to be the one, up to canonical transformations, described 
by Eqs.~(14) and (15).

\bigskip
\bigskip

\centerline{\normalsize {4. APPLICATION TO PERIODIC CHARTS}}

\bigskip

To close this paper, we briefly discuss the use of the group SO($4, 2$) for 
the classification of chemical elements. The importance of SO($4, 2$) for 
the periodic charts of neutral atoms and ions was noted by Barut \cite{Bar72-27} 
during the Rutherford Centennial symposium organised by Wybourne at the University 
of Canterbury (New Zealand) in 1971 and, independently, by Rumer and Fet 
\cite{RumFet72-28} (see also [52, 53]) in the former Soviet Union in 1971 too. Later, 
Byakov {\it et al}. \cite{3pre76-29}
further developed this group-theoretical approach of the periodic chart of chemical 
elements by introducing the direct product SO($4, 2$)$\otimes$SU(2). One of the deep 
reasons for using SO($4, 2$) relies upon the fact that the Madelung rule inherent to 
the {\it Aufbau Prinzip} can be described by the representation $h$ of SO($4, 2$) 
[50, 55, 56].

In the {\it \`a la} SO($4, 2$) or SO($4, 2$)$\otimes$SU(2) 
approach to the periodic table, the infinite-dimensional UIR 
$h$ of SO($4, 2$) is used for describing 
neutral atoms. Each atom thus appears as a particular 
partner for the representation $h\otimes[2]$ of the group 
SO($4, 2$)$\otimes$SU(2) where $[2]$ stands for the 
fundamental represention of SU(2). In fact, it is 
possible to connect two partners of the representation 
$h\otimes[2]$ by making use of shift operators of the Lie 
algebra of SO($4, 2$)$\otimes$SU(2). In other words, it is 
possible to pass from one atom to another one by means of 
raising or lowering operators. We thus obtain a chart with 
rows and columns for which the $n$th row contains $2n^2$ 
elements and the $l$th column contains an infinite number 
of elements. A given column corresponds to a family of chemical 
analogs in the standard periodic table and a given row may 
contain several periods of the standard periodic table.
Three features of the SO($4, 2$)$\otimes$SU(2) 
periodic table are the following: (i) hydrogen is in the 
family of the alkali metals, (ii) helium belongs to the 
family of the alkaline earth metals, and (iii) the inner 
transition series (lanthanides and actinides) as well as 
the transition series (iron group, palladium group and 
platinum group) occupy a natural place in the table. This 
is to be contrasted with the conventional tables with 8(9) 
or 18 columns where: (i) hydrogen is sometimes located in 
the family of the halogens, (ii) helium generally belongs 
to the family of the noble gases, and (iii) the lanthanide 
series and the actinide series are generally treated as 
appendages. The reader may consult Ref. [56] for more 
details.

To date, the use of SO($4, 2$) or SO($4, 2$)$\otimes$SU(2) in 
connection with periodic charts has been limited to 
qualitative aspects only, viz., classification of neutral 
atoms and ions as well. We would like to give here the main 
lines of a programme under development (inherited from nuclear physics and 
particle physics) for dealing with quantitative aspects.

The first step concerns the mathematics of the programme. The 
direct product group SO($4, 2$)$\otimes$SU(2) 
is a Lie group of order eighteen. Let 
us first consider the SO($4, 2$) part which is a semi-simple 
Lie group of order $r = 15$ and of rank $l = 3$. It has thus 
fifteen generators involving three Cartan generators (i.e., generators 
commuting between themselves). Furthermore, it has three invariant 
operators or Casimir operators (i.e., independent polynomials, 
in the enveloping algebra of the Lie algebra of SO($4, 2$), that 
commute with all generators of the group SO($4, 2$)). Therefore, 
we have a set of six ($3 + 3$) operators that commute between 
themselves: the three Cartan generators and the three Casimir operators. 
Indeed, this set is not complete from the mathematical point of 
view. In other words, the eigenvalues of the six above-mentioned 
operators are not sufficient for labelling the state vectors in 
the representation space of SO($4, 2$). According to a (not very 
well-known) result popularised by Racah, we need to find
$$
\frac{1}{2}(r - 3l) = 3
\eqno (29)
$$
additional operators in order to complete the set of the six preceding 
operators. This yields a complete set of nine ($6 + 3$) commuting operators 
and this solves the state labelling problem for the group SO($4, 2$). 
The consideration of the group SU(2) is trivial: SU(2) is a semi-simple Lie 
group of order $r = 3$ and of rank $l = 1$ so that $(r - 3l)/2 = 0$ 
in that case. As a result, we end up with a complete set of eleven ($9 + 2$) 
commuting operators. It is to be stressed that this result constitutes the key 
and original starting point of the programme. 

The second step establishes contact with chemical physics. Each of the eleven  
operators can be taken to be self-adjoint and thus, from the quantum-mechanical 
point of view, can describe an observable. Indeed, four of the eleven operators, 
namely, the three Casimir operators of SO($4, 2$) and the Casimir operator of SU(2),
serve for labelling the representation $h\otimes[2]$ of SO($4, 2$)$\otimes$SU(2) 
for which the various chemical elements are partners. The seven remaining operators 
can thus be used for describing chemical and physical properties of the elements
like:

\medskip

-ionization energies, 

-electron affinities, 

-electronegativities, 

-melting and boiling points, 

-specific heats, 

-atomic radii, 

-atomic volumes, 

-densities, 

-magnetic susceptibilities, 

-solubilities, etc. 

\medskip
 
\noindent In most cases, this can be done by expressing a chemical observable 
associated with a given property in terms of the seven operators which serve as 
an integrity basis for the various observables. Each observable can be developed 
as a linear combination of operators constructed from the integrity basis. This is 
reminiscent of group-theoretical techniques used in nuclear and atomic 
spectroscopy (cf. the Interacting Boson Model) or in hadronic spectroscopy (cf. the
Gell-Mann/Okubo mass formulas for baryons and mesons). 

The last step is to proceed to a diagonalisation process and then to 
fit the various linear combinations to experimental data. This 
can be achieved through fitting procedures concerning either a period of elements 
(taken along a same line of the periodic table) or a 
family of elements (taken along a same column of the 
periodic table). For each property this will lead to 
a formula or phenomenological law that can 
be used in turn for making predictions concerning the chemical elements for 
which no data are available. In addition, it is hoped that this will shed 
light on the new patterns of the periodic 
table recently discovered; along
this vein, the singularity principle, the diagonal relationships and the inert
pair effect [57-59] are patterns that can be described on a 
mathematical basis by using ladder operators. 

This programme, referred to as the KGR programme, was briefly 
presented at the 2003 Harry Wiener International Conference 
\cite{Kib04-31}. It is presently under progress. The concept of
$q$-deformation (as arising in the theory of quantum groups), 
already applied to the classification of atoms and ions 
[61, 62], is of potential interest for this 
programme. Finally, it should be emphasized that this programme might be adapted 
to molecules since subgroups of SO($4, 2$) were successfully used 
for periodic systems of molecules [63, 64]. 

\baselineskip = 0.55 true cm

\end{document}